\documentclass[showpacs,showkeys,aps,prb,twocolumn, superscriptaddress]{revtex4-1}

\usepackage[sort&compress]{natbib}
\usepackage{graphics}
\usepackage{dcolumn}
\usepackage{bm}
\usepackage{amsmath,empheq}
\usepackage{setspace}
\usepackage{hyperref}

\begin{document}

\title{Frustrated magnetism in tetragonal CoSe, analogue to superconducting FeSe} 

\author{Brandon Wilfong}
\affiliation{Department of Chemistry and Biochemistry, University of Maryland, College Park, Maryland 20742, United States}
\affiliation{Center for Nanophysics and Advanced Materials, University of Maryland, College Park, Maryland 20742, United States}
\author{Xiuquan Zhou}
\affiliation{Department of Chemistry and Biochemistry, University of Maryland, College Park, Maryland 20742, United States}
\author{Hector Vivanco}
\affiliation{Department of Chemistry and Biochemistry, University of Maryland, College Park, Maryland 20742, United States}
\author{Daniel  J. Campbell}
\affiliation{Center for Nanophysics and Advanced Materials, University of Maryland, College Park, Maryland 20742, United States}
\affiliation{Department of Physics, University of Maryland, College Park, Maryland 20742, United States}
\author{Kefeng Wang}
\affiliation{Center for Nanophysics and Advanced Materials, University of Maryland, College Park, Maryland 20742, United States}
\affiliation{Department of Physics, University of Maryland, College Park, Maryland 20742, United States}
\author{Dave Graf}
\affiliation{National High Magnetic Field Laboratory, 1800 East Paul Dirac Drive, Tallahassee, Florida 32310, USA}
\author{Johnpierre Paglione}
\affiliation{Center for Nanophysics and Advanced Materials, University of Maryland, College Park, Maryland 20742, United States}
\affiliation{Department of Physics, University of Maryland, College Park, Maryland 20742, United States}
\author{Efrain Rodriguez}
\email{efrain@umd.edu}
\affiliation{Department of Chemistry and Biochemistry, University of Maryland, College Park, Maryland 20742, United States}
\affiliation{Center for Nanophysics and Advanced Materials, University of Maryland, College Park, Maryland 20742, United States}

%%%%%%%%%%%%%%%%%%%%%%%%%%%%%%%%%%%%%%%%%%%%%%%%%%%%%%%%%%%%%%%%%%%%%
%% Abstract
%%%%%%%%%%%%%%%%%%%%%%%%%%%%%%%%%%%%%%%%%%%%%%%%%%%%%%%%%%%%%%%%%%%%%

\begin{abstract}
Recently synthesized metastable tetragonal CoSe, isostructural to the FeSe superconductor, offers a new avenue for investigating systems in close proximity to the iron-based superconductors. We present magnetic and transport property measurements on powders and single crystals of CoSe. High field magnetic susceptibility measurements indicate a suppression of the previously reported 10 K ferromagnetic transition with the magnetic susceptibility, exhibiting time-dependence below the proposed transition. Dynamic scaling analysis of the time-dependence yields a critical relaxation time of $\tau^{*} = 0.064 \pm 0.008 $ s which in turn yields activation energy E$_{a}^{*}$ = 14.84 $\pm$ 0.59 K and an ideal glass temperature T$_{0}^{*}$ = 8.91 $\pm$ 0.09 K from Vogel-Fulcher analysis. No transition is observed in resistivity and specific heat measurements, but both measurements indicate that CoSe is metallic. These results are interpreted on the basis of CoSe exhibiting frustrated magnetic ordering arising from competing magnetic interactions. Arrott analysis of single crystal magnetic susceptibility has indicated the transition temperature occurs in close proximity to previous reports and that the magnetic moment lies solely in the $ab$-plane. The results have implications for understanding the relationship between magnetism and transport properties in the iron chalcogenide superconductors.
\end{abstract}

\maketitle

%%%%%%%%%%%%%%%%%%%%%%%%%%%%%%%%%%%%%%%%%%%%%%%%%%%%%%%%%%%%%%%%%%%%%
%% Introduction
%%%%%%%%%%%%%%%%%%%%%%%%%%%%%%%%%%%%%%%%%%%%%%%%%%%%%%%%%%%%%%%%%%%%%

\section{Introduction}

\par  The iron-based superconducters are composed of Fe$^{2+}$ square lattices stacked to form layered materials. For example, the simple FeSe superconductor contains stacked layers of Fe$^{2+}$ centers tetrahedrally-coordinated to selenide anions. Remarkably, its $T_c$ of 8 K, \cite{hsu2008superconductivity} can be increased to 65 - 100 K when isolated as a single layer. \cite{ge2015superconductivity, he2013phase} Therefore, it is the square sublattice of $d$-cations that may hold the key to understanding the physical properties of these systems. In this article, we have completely replaced the Fe$^{2+}$ cations in FeSe with Co$^{2+}$, and studied its magnetization, magnetotransport and specific heat properties to further explore the physics of metal square lattices.

In addition to crystal structure, the relationship between magnetism and superconductivity is of paramount importance for these layered chalcogenides. In the iron pnictide superconductors (e.g. BaFe$_{2}$As$_{2}$ and LaOFeAS), suppression of the parent antiferromagnetic (AFM) phase can lead to the emergence of superconductivity.\cite{ishida2009extent, wen2011materials} However, no long-range magnetic ordering has been observed in any of the FeSe or FeS superconductors. Although antiferromagnetism was found in Fe$_{1+x}$Te, the origin of its magnetism is different from that of the pnictides, and it is largely influenced by the amount of interstitial iron. \cite{FeTe_Rodriguez, Rodriguez_FeTe, Stock_FeTe} Thus, it is less clear how magnetism and superconductivity interact in the Fe\textit{Ch} (\textit{Ch} = chalcogenide) systems compared to their pnictide counterparts. 

\begin{figure}[t!]
	\includegraphics[width=3in, height=2in]{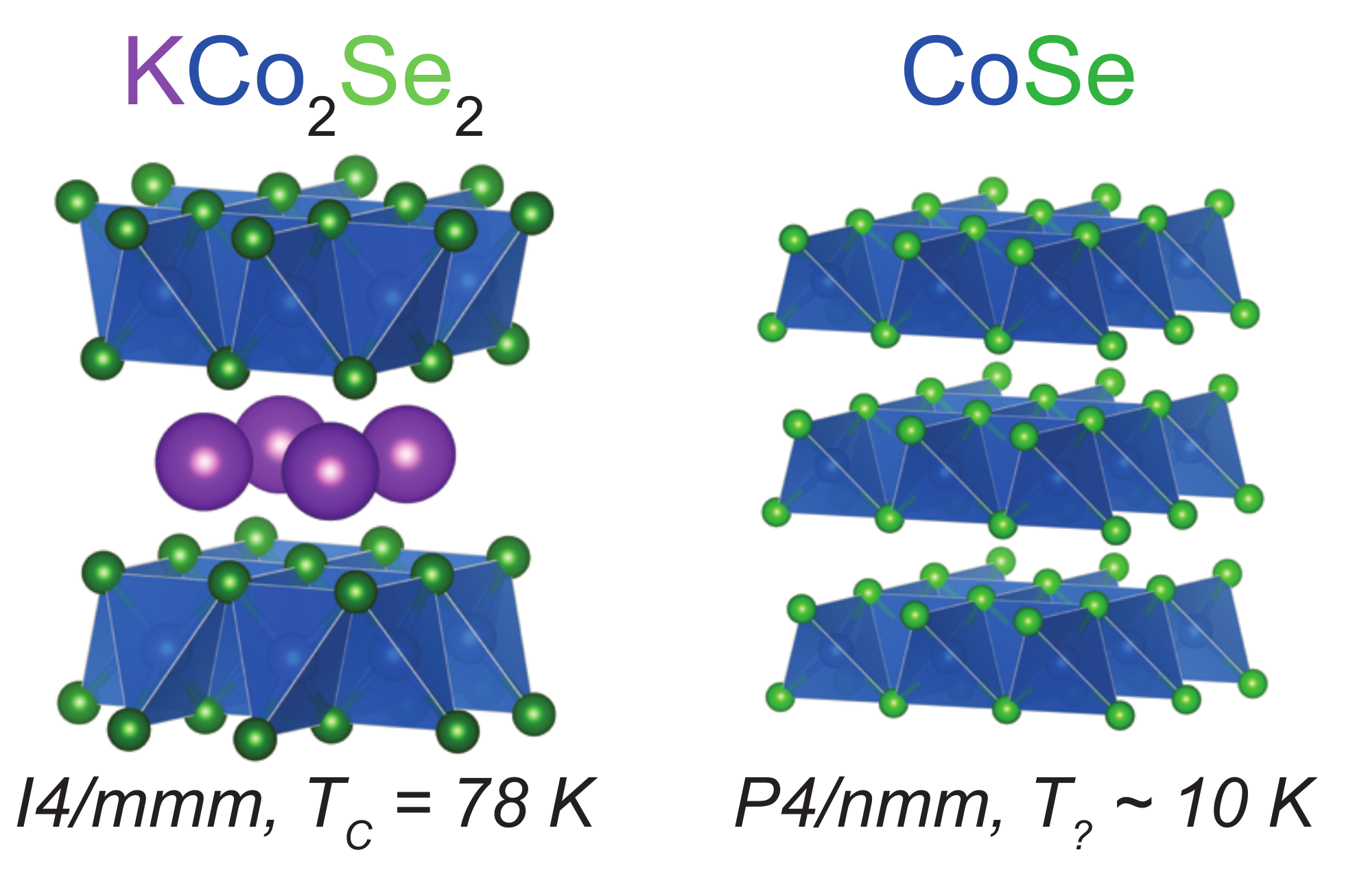}
 	\caption{Crystal structures of KCo$_{2}$Se$_{2}$ and CoSe. }
  	\label{figure:CoSe_structure}
\end{figure}

Currently, one key issue is that isostructural systems to Fe\textit{Ch} are limited due to synthetic challenges.  Previously, we have overcome this challenge by topochemical means to convert KFe$_{2}$S$_{2}$ to superconducting FeS.\cite{BorgFeS} Using a similar method, we successfully prepared two new Fe\textit{Ch} analogues, tetragonal CoSe and CoS. \cite{ZhouJACS}  The ferromagnetic ordering from 78 K in KCo$_{2}$Se$_{2}$\cite{Greenblatt_AM2X2} to 10 K in CoSe\cite{ZhouJACS} was suppressed by de-intercalation of potassium cations to form pure CoSe as shown in Figure \ref{figure:CoSe_structure}. These new Co-based phases are promising for understanding the Fe-based superconductors due to their structural and electronic proximity.

%-----------------------------------------------------------------------------------
%-----------------------------------------------------------------------------------

\par Much of the work performed to understand the magnetism in iron pnictides has been done with those that adopt the ThCr$_{2}$Si$_{2}$ structure-type (``122''-system). This structure-type allows for a wider range of substitutions on the metal, anion and interlayer cation sites to study doping effects.\cite{Zhang_122Pn, paglione2010high, Zhou_a, Zhou_b} There has been extensive work on the cobalt analogues to ``122'' iron pnictides, \textit{A}Co$_{2}$\textit{Pn}$_2$, with various interlayer alkali or alkali-earth cations (\textit{A}).\cite{Quirinale_CaCo2As2, Anand_CaCo2As2, Jayasekara_SrCo2As2, pandey2013crystallographic, Anand_BaCo2As2, Mansart_ACo2As2, Kovnir2011a, Tan2016}  The observed magnetism in these pnictides was largely tuned by size and electronic effects from changing the Co\textit{Pn} layer distances. 

\par An intriguing question is: can CoSe be tuned into a superconductor like FeSe? By directly comparing their band structures, CoSe chould share similar electron-hole pockets with FeSe if the electron filling level is reduced. \cite{ZhouJACS} Therefore, it may be possible to tune CoSe into a superconductor by increasing the Co oxidation state to form $d^6$ cations isoelectronic to Fe$^{2+}$. In order to investigate this, two fundamental factors must be understood: 1) the character of the magnetic interactions within the Co square lattice, and 2) how its magnetism compares to other Fe\textit{Ch} based superconductors.  Here, we have performed extended magnetic and transport characterizations to understand the magnetism within CoSe and its proximity to superconductivity in related FeSe.

%%%%%%%%%%%%%%%%%%%%%%%%%%%%%%%%%%%%%%%%%%%%%%%%%%%%%%%%%%%%%%%%%%%%%
%% Experimental
%%%%%%%%%%%%%%%%%%%%%%%%%%%%%%%%%%%%%%%%%%%%%%%%%%%%%%%%%%%%%%%%%%%%%

\section{Experimental Methods}

\par Single crystals and powders of CoSe were synthesized following the previous method in literature. \cite{ZhouJACS} Crystals of CoSe were lustrous silver with high degree of layered morphology. 

\par  Temperature dependent DC (direct current) magnetic susceptibility measurements were carried out using a Quantum Design Magnetic Susceptibility Measurement System (MPMS) on powder samples of tetragonal CoSe. Field-cooled (FC) and zero field-cooled (ZFC)  measurements were taken from 1.8 K to 300 K with various applied magnetic field strengths. Magnetic hysteresis measurements were carried out using a  PPMS DynaCool utilizing a vibrating sample magnetometer (VSM) taken at a series of temperatures with applied magnetic field between $H = \pm 14$ T on single crystals of CoSe mounted on a quartz paddle via Ge 7031 varnish.

\par AC (alternating current) magnetic susceptibility was measured with a 14 T Quantum Design Physical Property Measurement System (PPMS-14) on powder samples of tetragonal CoSe. Zero field-cooled measurements were taken from 35 K to 1.8 K with an AC-field of 10 Oe and AC-frequencies of 10 Hz to 10 kHz. Due to the instrument setup, a residual DC field within the PPMS-14 ranged from 40 Oe to 100 Oe. 

\par Electrical transport measurements were preformed using a 9 T Quantum Design Physical Property Measurement System (PPMS-9) with single crystals of CoSe mounted on a Quantum Design AC transport puck. Electrical resistivity was measured using the four-probe method with gold wire and contacts made with silver paste. The temperature and field dependence of longitudinal electrical resistivity was measured in a range from 300 K to 1.8 K with applied fields up to 9 T. 

\par Electrical transport measurements at fields up to 31 T were performed at the DC Field Facility of the National High Magnetic Field Laboratory in Tallahassee, Florida. Angular dependence measurements at base temperature of the He-3 system (500-600 mK) were made by rotating the sample plane ($ab$-plane) from perpendicular (0 degrees) to parallel (90 degrees) to the applied field. Temperature dependent magnetotransport was measured for applied field both perpendicular and parallel to the sample plane between base temperature and 12 K.

\par Heat capacity measurements were preformed using the PPMS-14. Heat capacity measurements on tetragonal CoSe single crystals  yielded poor results due to low thermal contacts arising from the micaceous nature of the CoSe flakes. Consequently, a pressed pellet of CoSe ground single crystals was used for the heat capacity measurements performed with the relaxation technique. \cite{QuantumDesign, HwangHC, BachmannHC}

\par All density functional theory (DFT) \cite{Hohenberg, Kohn} calculations were performed by using the Vienna Ab-initio Simulation Package (VASP)\cite{KresseThesis, KresseMD, KresseCalc, KresseIterative} software package with potentials using the projector augmented wave (PAW)\cite{BlochlPAW} method. The exchange and correlation functional were treated by the generalized gradient approximation (PBE-GGA).\cite{PerdewGradient}  The cut-off energy, 450 eV, was applied to the valance electronic wave functions expanded in a plane-wave basis set. A Monkhorst-Pack\cite{MonkhorstBrillouin} generated 23$\times$23$\times$17 k-point grid was used for the Brillouin-zone integration to obtain accurate electronic structures.

%%%%%%%%%%%%%%%%%%%%%%%%%%%%%%%%%%%%%%%%%%%%%%%%%%%%%%%%%%%%%%%%%%%%%
%% Results
%%%%%%%%%%%%%%%%%%%%%%%%%%%%%%%%%%%%%%%%%%%%%%%%%%%%%%%%%%%%%%%%%%%%

\section{Results}

%%%%%%%%%%%%%%%%%%%%%%%%%%%%
%% Magnetic measurements and properties
%%%%%%%%%%%%%%%%%%%%%%%%%%%%

\subsection{Magnetic properties}
Our previous work demonstrated the suppression of ferromagentism from 78 K in KCo$_{2}$Se$_{2}$\cite{Greenblatt_AM2X2} to 10 K in CoSe\cite{ZhouJACS}. However, due to the very low ordering moment as well as the proximity to the iron-based superconductors, a more detailed investigation of the magnetism and electronic properties has been undertaken. 

\par
\begin{figure}[t!]
	\includegraphics[width=3in, height=2.75in]{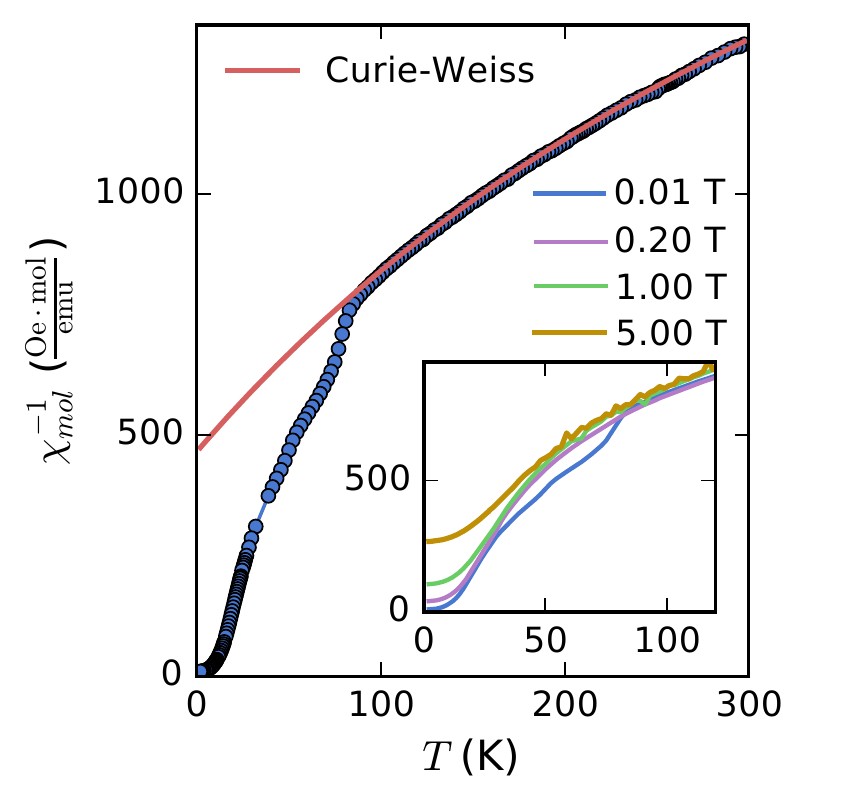}
 	\caption{Inverse magnetic susceptibility of CoSe \textit{vs.} temperature measured in applied field of 100 Oe. The inverse magnetic susceptibility is fit from 100 K to 300 K to the Curie-Weiss law plus a temperature-independent term. The inset shows inverse magnetic susceptibilities for different applied DC fields (0.2 T, 1 T and 5 T) to emphasize the change in slope near 82 K. }
  	\label{figure:CoSe_CurieWeiss}
\end{figure}

  Figure \ref{figure:CoSe_CurieWeiss} shows the temperature dependence of inverse FC magnetic susceptibility for ground single crystal samples of CoSe at various applied DC fields. The inverse susceptibility was fit in the paramagnetic range from 100 K to 300 K to the Curie-Weiss law:

\begin{equation}
\chi_{mol} = \chi_{0} + \frac{C}{T - \Theta_{CW}}
\end{equation}

\noindent where $\chi_{0} = 3.52 \times10^{-4} \frac{emu}{Oe \cdot mol} $ accounts for parasitic paramagnetic and diamagnetic contributions, $C = 0.1579 \frac{emu\cdot K}{Oe\cdot mol}$ denotes the Curie constant, and $\Theta_{CW}$ = -87.29 K is the Weiss constant. A strongly negative Weiss constant empirically indicates predominant antiferromagnetic fluctuations. 

%-----------------------------------------------------------------------------------
%-----------------------------------------------------------------------------------

\par The frustration parameter, $f$, for a magnetic system is defined as the ratio of the absolute value of the Weiss constant and the observed ordering temperature from magnetic susceptibility:\cite{balents2010spin}

\begin{equation}
f = \frac{\mid\Theta_{CW}\mid}{T_{C}}
\end{equation}

\noindent We obtain a frustration parameter of approximately 8.7, indicating strong suppression the magnetic ordering temperature. The inset of Figure \ref{figure:CoSe_CurieWeiss} displays the inverse susceptibility behavior with different applied fields; it is shown that the paramagnetic regime ($> 100$ K) does not change, but the deviation at approximately 82 K shows differing behavior with applied DC field. Empirically, in the frustrated regime ($T_{c} < T < \lvert\Theta_{CW}\lvert$) increasing field drives the system toward increasing antiferromagnetic fluctuations as the slope of $\chi^{-1}(T)$ decreases.

\par
\begin{figure}[h]
	\includegraphics[width=3in, height=3in]{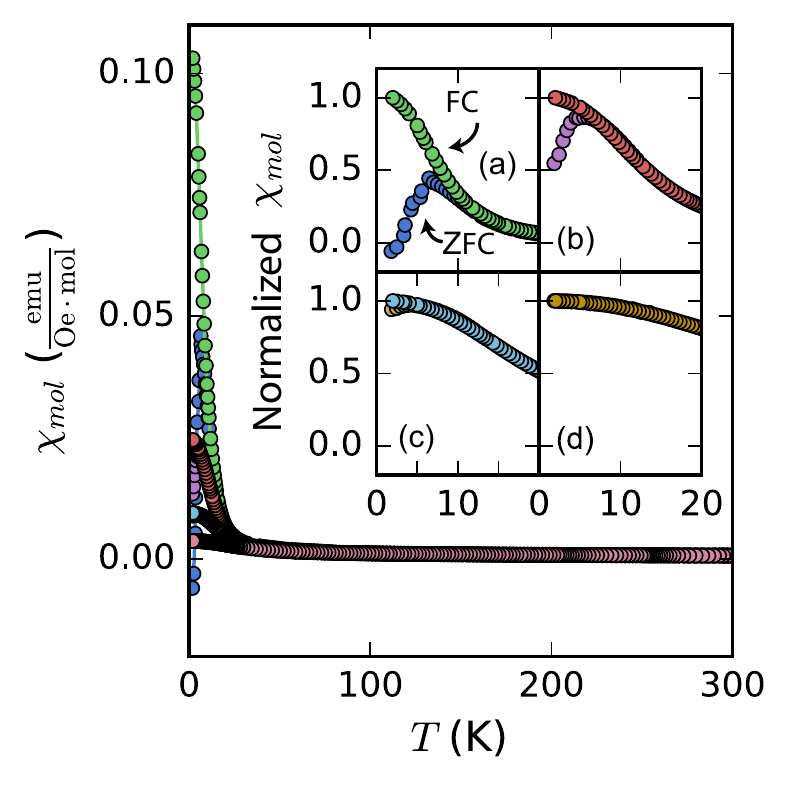}
 	\caption{Magnetic susceptibility of CoSe \textit{vs.} temperature measured in various applied fields. The insets show the zoomed region close to the transition temperature; ZFC (Zero field-cooled) and FC (field-cooled) curves are shown by arrows which indicate the irreversibility of the magnetic ordering in the system at low fields. The bifurcation of ZFC-FC curves at low applied field  (a) = 0.01 T and (b) = 0.2 T is destroyed with high applied fields (c) = 1 T and (d) = 5 T turning the system into a paramagnetic state with no irreversibility.}
  	\label{figure:CoSe_XvT}
\end{figure}

 Our earlier work showed that the magnetic susceptibility of CoSe exhibited a ferromagnetic transition at 10 K, but the discontinuity at 10 K was not a classic example of a ferromagnetic transition. In order to explore this, the temperature dependence of magnetic susceptibility was measured at different fields to see how the transition was altered. Insets of Figure \ref{figure:CoSe_XvT} shows the magnetic susceptibility from 10 K to 1.8 K at various applied fields. At low fields, 0.01 T, the transition at 10 K is clear from the bifurcation of the zero-field cooled (ZFC) and field-cooled (FC). However, as the field is increased, the transition temperature is suppressed until at high fields, $ > 2 $ T, there is a complete suppression of the ZFC-FC splitting indicative of complete suppression of long-range ferromagnetic ordering.

%-----------------------------------------------------------------------------------
%-----------------------------------------------------------------------------------

\par The closing of the normal ZFC-FC splitting at the proposed ferromagnetic transition is a hallmark of spin glass behavior as opposed to classic ferromagnetism. \cite{mydosh1993spin} Without a sufficiently applied field, spins are able to ``freeze" in the random orientation of spin glass yielding net magnetization opposing the applied field in the ZFC process. With a stronger field, the ``freezing" is destroyed as the spins are forced to align with the applied field. We can rule out superparamagnetism as a possible explanation as we have observed remanent magnetization and magnetic hysteresis for CoSe which would not occur in a superparamagnetic material. \cite{ZhouJACS}  In order to observe the glassy character in CoSe, we performed AC magnetic susceptibility measurements to probe the time-dependence of the magnetization around the transition temperature.

%-----------------------------------------------------------------------------------
\begin{figure}[t!]
	\includegraphics[width=3in, height=6.75in]{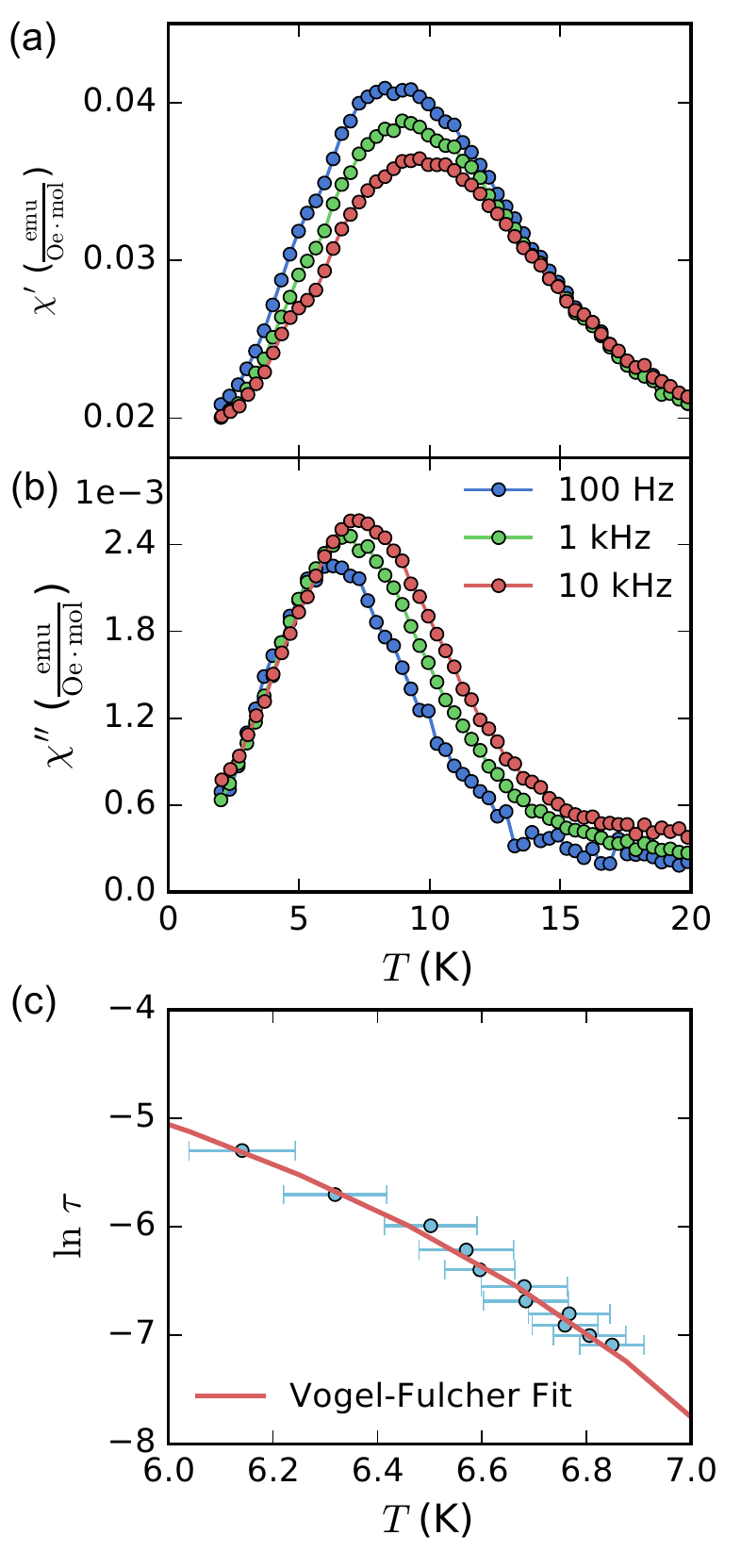}
 	\caption{AC magnetic susceptibility measured with various driving frequencies. The applied AC field was 10 Oe and the residual DC applied field due to internal instrumentation was 40 Oe to 100 Oe. (a) The real parts of magnetic susceptibility ($\chi'$) and (b) the imaginary parts ($\chi''$) parts. (c) Temperature dependence of $\chi"$ peaks at various driving frequencies (100 to 1200 Hz) and a fit with the Volger-Fulcher law.}
  	\label{figure:CoSe_ACMS}
\end{figure}
%-----------------------------------------------------------------------------------

\par AC-susceptibility measurements use an applied field with a time-dependent waveform to produce a time-dependent response in the material. It can therefore probe spins fluctuating with time such as in spin glasses or strongly frustrated systems.\cite{mydosh1993spin} Figure \ref{figure:CoSe_ACMS} shows the real ($\chi'$) and imaginary ($\chi''$) parts of magnetic susceptibility as a function of temperature near the transition. Frequency dependence in $\chi'$ appears below 10 K, and accompanying non-zero peaks in $\chi''$ indicate some out-of-phase contributions to the magnetic susceptibility. Thus, time-dependence in the magnetic domain size arises below 10 K, and any mangetic ordering appears dynamic down to base temperature.

%%%%%%%%%%%%%%%%%%%%%%%%%%%%%%%%%%%%%%%%
%\par A figure of merit used to describe systems in which the magnetic susceptibility exhibits frequency dependence is the frequency sensitivity, $\kappa$\%cite{mydosh1993spin, mydosh1996disordered}
%
%\begin{equation}
%kappa = \frac{\Delta T_{f}}{T_{f}\Delta log(\omega)}
%\end{equation}

%The frequency sensitivity, $\kappa = 0.09(9)$, calculated from the change in $\chi'$ peaks as a function of frequency, places CoSe into the class of spin-%glass-like materials as the freezing temperature shows more dependence on frequency than canonical spin glasses ($10^{-3} - 10^{-2}$), but less than %superparamagnetic materials ($10^{-1} - 10^{0}$). \cite{mydosh1993spin} 
%%%%%%%%%%%%%%%%%%%%%%%%%%%%%%%%%%%%%%%%

\par A fit to the non-zero $\chi''$ peaks with the Arrhenius law would be simple yet inadequate for canonical spin glasses and spin-glass-like materials.  The transition into the glassy state is more than a simple thermal activation process, and magnetic moments can also be strongly interacting.\cite{mydosh1993spin} A more phenomenological approach that incorporates different regimes of coupling above and within the glassy state uses the Vogel-Fulcher law:\cite{mydosh1993spin, binder1986spin}

\begin{equation}
\tau = \tau_{0}\cdot\mathrm{exp}\bigg(\frac{E_{a}}{k_{B}(T_f-T_{0})}\bigg)
\end{equation}

\noindent where $T_f$ is the temperature of the $\chi''$ peaks, $\tau_{0} = 1/\omega_{0}$ is the characteristic relaxation time, and $k_B$ the Boltzmann constant. The added parameter, $T_{0}$, describes the `ideal glass temperature' where the coupling of the system effectively changes to give rise to new phenomena. \cite{binder1986spin,shtrikman1981theory, mydosh1993spin} 

\par Our modelling of the AC susceptibility data with the Vogel-Fulcher law is shown in Figure \ref{figure:CoSe_ACMS}c.  The temperature values for  $T_{f}$ were fit by Gaussian curves in the range from 5-12 K. The fit yields parameters: $\tau_{0} = 0.67 \pm 1.61 $ s, $E_{a} = 12.75 \pm 10.47 $ K, and $T_{0} = 8.74 \pm 0.89$ K. The large degree of uncertainty in the relaxation time and activation energy comes from the high correlation between $\tau_{0}$ and $T_{0}$ parameters and narrow temperature range of the $\chi''$ peaks.

\par The lack of meaningful values from the initial Vogel-Fulcher fit led us to perform additional analysis using a dynamical scaling model. Dynamical scaling relates the relaxation time of an observable to a correlation length that scales with a power law near the transition temperature. We consider scaling of the frequency-dependent transition temperature from the $\chi''$ peaks such that: \cite{TholenceScaling}

\begin{equation}
\tau = \tau^{*}\bigg(\frac{T_{c} -  T_{f}}{T_{f}}\bigg)^{-zv}
\end{equation}

\noindent where $T_c$ is the critical temperature, $\tau^{*}$ the critical relaxation time, and $zv$ the critical exponent. Our fit yields $\tau^{*} = 0.064 \,\pm\, 0.008 $ and $zv = 5.47 \pm 0.21$, which fall into the general range of spin-glass and glassy-like materials \cite{TholenceScaling}. 

\par Substituting the value of $\tau^{*}$ for the the characterstic relaxation time in the Vogel-Fulcher law, we obtain more precise values for the activation energy ($E_{a}^{*} = 14.84 \,\pm\, 0.59$ K) and ideal glass temperature ($T_{0}^{*} = 8.91 \pm 0.09$ K). The obtained critical relaxation temperature is significantly higher than canonical spin-glass materials, but compatible with Monte-Carlo modeling of a 3D Ising spin glass. \cite{OgielskiIsing, OgielskiIsing2}

\subsection{Transport Properties}

\begin{figure}[t!]
	\includegraphics[width=3.in, height=6.75in]{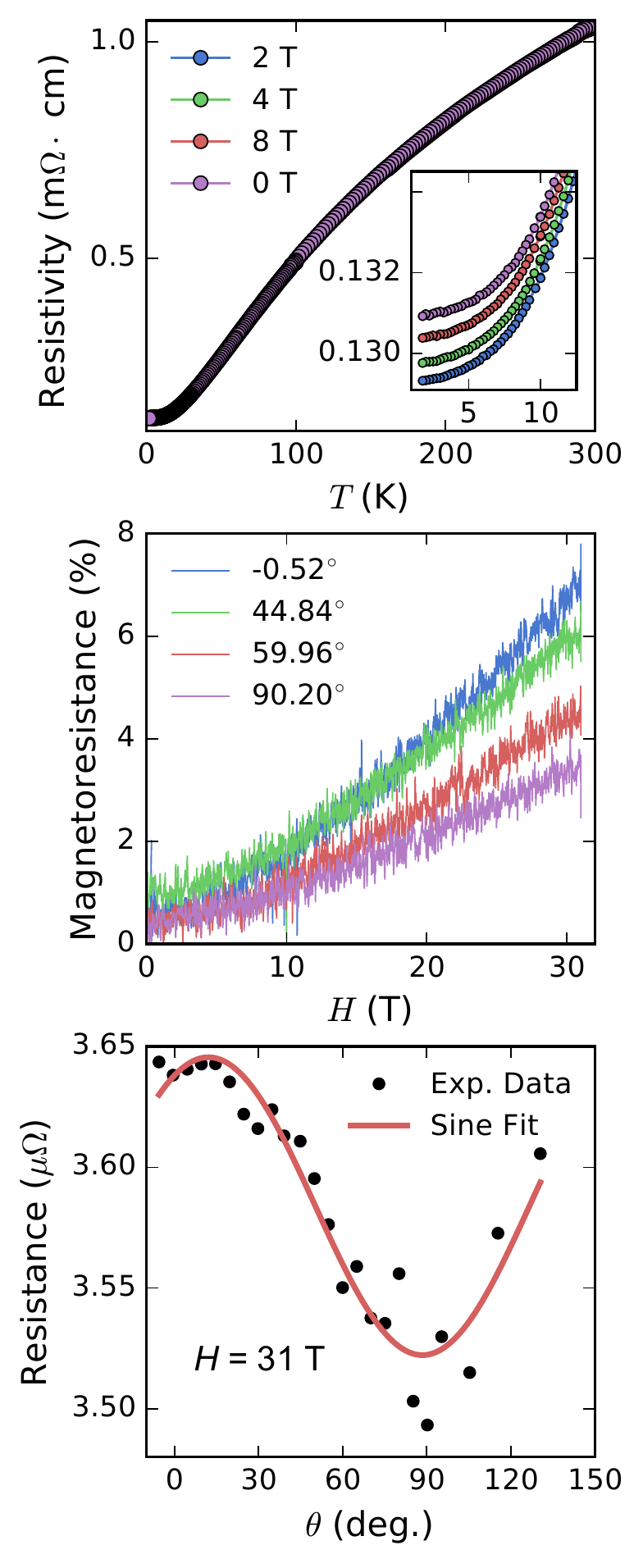}
 	\caption{Electrical transport measurements of CoSe single crystals obtained through de-intercalation of KCo$_{2}$Se$_{2}$. (a) Temperature dependence of longitudinal resistivity at various applied fields with inset around the transition temperature. (b) Normalized longitudinal magnetoresistance up to 31 T with different applied field directions by sample rotation. (c) Angular dependence of longitudinal magnetoresistance at an applied field of 31 T.  The magnetoresistance is fit with a sinusoidal dependence to the field angle.}
  	\label{figure:CoSe_Resistance}
\end{figure}

To further probe the dynamics of the transition within CoSe, we have employed more electronic transport  measurements. Figure \ref{figure:CoSe_Resistance}a shows the temperature dependence of electrical resistivity for CoSe. For a truly ordered material, one would anticipate a noticeable change in the resistivity near the critical point. However, no such anomaly occurs in the resistivity measurements. This lack of an anomaly could be understood on the basis of weak ferromagnetism as the observed moment of CoSe via neutron diffraction is very small. \cite{ZhouJACS} 

%-----------------------------------------------------------------------------------
%-----------------------------------------------------------------------------------

\par We observe positive magnetoresistance for all applied field directions (Figure \ref{figure:CoSe_Resistance}b), which does not occur in typical ferromagnets. The positive magnetoresistance can be interpreted in two ways: 1) the spins have no fixed direction and are randomly distributed as would be the case for a glass-like material, or 2) the spins are fixed but their associated moments are so small that their contribution to scattering is negligible. 

\par The complete angular dependence of the resistance versus field direction at 31 T (Figure  \ref{figure:CoSe_Resistance}c) shows two-fold symmetry, which is due to the geometry of the four-probe longitudinal measurements. Angular measurements in other planes are not possible due to sample morphology. CoSe crystals are highly layered and micaecous so that only allow the $ab$-plane is available as the wiring surface. 

\par
\begin{figure}[t!]
	\includegraphics[width=3in, height=3in]{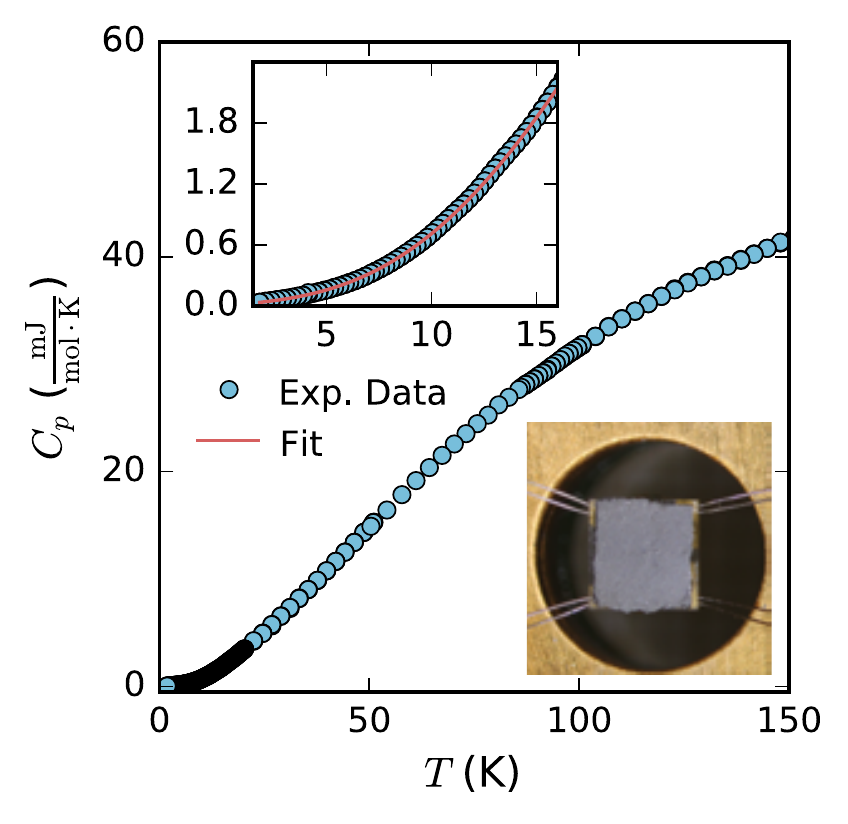}
 	\caption{Temperature dependenct specific heat of a pressed pellet of CoSe from 150 K to base temperature. Upper inset shows the temperature dependence near the transition temperature as well as a fit to a specific heat model accounting for electronic and vibrational components in the range 1.8 K to 15 K.}
  	\label{figure:CoSe_HC}
\end{figure}

 We performed specific heat measurements from 1.8 - 150 K (Figure \ref{figure:CoSe_HC}) on a pressed pellet of CoSe obtained through the potassium de-intercalation route. The micaecous nature of the single crystals caused poor thermal coupling between the sample and the heating platform, and we therefore utilized a pressed pellet of CoSe. For comparision, we also performed specific heat measurements of KCo$_{2}$Se$_{2}$ single crsytals,  known from previous studies to exhibit a clear ferromagnetic transition below 80 K.\cite{yang2013magnetic,Greenblatt_AM2X2} Our own heat capacity measurements of KCo$_{2}$Se$_{2}$ confirm a clear transition at 78 K.\cite{supp} The temperature dependence of the specific heat for CoSe, however, shows no anomaly near 78 K.

%-----------------------------------------------------------------------------------
%-----------------------------------------------------------------------------------

\par The inset of Figure \ref{figure:CoSe_HC} shows a zoomed in region around the transition observed in magnetic susceptibility measurements with a fit to a conventional specific heat model. There is no apparent discontinuity in the specific heat in this region, which indicates a lack of a distinct phase transition. This result either supports a glass-like material,\cite{binder1986spin, mydosh1993spin} or that the magnetic ordering does not change the energy scale due to the low ordering moment of a weak itinerant ferromagnet.\cite{yelland2005ferromagnetic, bhattacharyya2011investigation} 

\par The low temperature region of the specific heat, T $< 15$ K, was fit to a general model to extract electronic and vibrational contributions. \cite{kittel2005introduction}

\begin{equation}
\mathrm{C}_{p} = \gamma T + \beta T^{3} + cT^{5}
\end{equation}

\noindent where the $\gamma$-term accounts for electronic contributions and $\beta/c$-terms for vibrational contributions. The fit yields a $\gamma$ = 15.7 mJ mol$^{-1}$ K$^{-2}$, significantly larger than in the iron-based analogues FeSe and FeS (5.4 and 5.1 mJ mol$^{-1}$ K$^{-2}$, respectively). \cite{BorgFeS, McQueenFeSe} This could indicate stronger electron correlations in the cobalt system. However, recent  angle-resolved photoemission spectroscopy (ARPES) work on related KCo$_{2}$Se$_{2}$ indicated weaker electron correlations in the cobalt system than in the KFe$_{2}$Se$_{2}$ analogue.\cite{liu2015orbital} A possible explanation for the larger $\gamma$ in CoSe than in FeSe is that it arises from spin fluctuations present in a weak intinerant ferromagnet.\cite{kaul1999spin, moriya1979recent} 

\par We can use the parameter $\beta = 6.2 \times 10^{-4}$  mJ mol$^{-1}$ K$^{-4}$ to calculate the Debye temperature, $\Theta_D$, for CoSe by the relation: \cite{kittel2005introduction} 

\begin{equation}
\Theta_{D} = \bigg(\frac{12\pi^{4}nR}{5\beta}\bigg)^{1/3}
\end{equation}

\noindent where $R$ is the universal gas constant. This fit yields a $\Theta_{D} = 232 $ K. We added the $T^{5}$ term since the $T^{3}$ contribution is generally only applicable up to $\Theta_{D}/50 = 4.6 $ K. \cite{tari2003specific} The resulting $c$ is $ -5.9 \times 10^{-7}$ mJ mol$^{-1}$ K$^{-6}$, two orders of magnitude lower than the iron analogue.

%%%%%%%%%%%%%%%%%%%%%%%%%%%%%%%%%%%%%%%%%%%%%%%%%%%%%%%%%%%%%%%%%%%%%
%% Discussion
%%%%%%%%%%%%%%%%%%%%%%%%%%%%%%%%%%%%%%%%%%%%%%%%%%%%%%%%%%%%%%%%%%%%%

\section{Discussion}	

\subsection{Ground state of CoSe}

Despite the structural simplicity of CoSe, its magnetic ground state is less straightforward. Initial temperature dependence of magnetic susceptibility indicated a ferromagnetic transition at 10 K corroborated by powder neutron diffraction work.\cite{ZhouJACS} When considering itinerant systems, it is often useful to evaluate Stoner's criterion for ferromagnetism in the system where the enhanced susceptibility $\chi_S$ is given by:\cite{mohn2006magnetism}

\begin{equation}
\chi_{S} = \frac{\chi_{P}}{1-F} = \frac{\chi_{P}}{1-D_{band}(E_{f})I_{s}/2}
\end{equation}

\noindent where $D_{band}(E_{f}$) is the density of states at the Fermi level, $I_{s}$ is the Stoner factor for Co ($\sim 0.9$ eV) divided by two to account for the two Co atoms per unit cell, and $\chi_P$ is Pauil paramagnetic susceptibility. The denominator allows us to formulate the Stoner's criterion such that  $F = D_{band}(E_{f})I_{s}/2$. We performed DFT calculations on CoSe which yielded $D_{band}(E_{f}) = 7.33 $ states/eV for non-spin dependent calculations. \cite{supp} Thus, this result leads to $F = 3.29 > 1 $ which indicates that CoSe should have a ferromagnetic ground state. 

\par The structurally related KCo$_{2}$Se$_{2}$ exhibits a ferromagnetic transition at approximately 78 K measured on single crystals. The previous work used anisotropic single crystal measurements of magnetization to show that the magnetic moment resides completely in the $ab$-plane for the CoSe layers. \cite{yang2013magnetic} No neutron diffraction work has been reported to date on KCo$_{2}$Se$_{2}$, but the fairly large spacing between CoSe layers and anisotropic susceptibility indicates that the moment is likely in the $ab$-plane. When we remove the interlayer potassium ions and reduce the CoSe interlayer spacing from $\sim$ 6.92 $\text{\AA}$ $ $ in KCo$_{2}$Se$_{2}$ to 5.33 $\text{\AA}$ in CoSe, we can consider how these adjacent planes may begin to interact. 

\par Figure \ref{figure:CoSe_CurieWeiss} showed that the magnetic susceptibility of CoSe displayed Curie-Weiss behavior above 100 K yielding a strongly negative Weiss constant, \mbox{$\Theta_{CW}~ = ~-87.29~$K}. Although CoSe is an itinerant electron system, we can minimally consider a square lattice Heisenberg model, \cite{pandey2013crystallographic, johnston2012heisenberg}  for which similar models have been applied extensively to the FeSe system, \cite{yu2015antiferroquadrupolar, ma2009first, wang2015nematicity, glasbrenner2015effect} to yield:

\begin{equation}
\Theta_{CW} = -\frac{(J_{1} + J_{2})}{k_{B}}
\end{equation}

\noindent where $J_1$ and $J_2$ describe the nearest-neighbor and next-nearest-neighbor interactions on the square lattice, respectively. In this case we see that $J_{1} + J_{2}$ = 87.29 K = 7.53 meV and that the exchanges should be antiferromagnetic based on the inverse susceptilibity data. At approximately 80 K,  $\chi^{-1}(T)$ increases its slope so that the Weiss field changes to a positive value, possibly indicative of increasingly ferromagnetic fluctuations in this lower temperature regime. 

\par Interestingly, specific heat, magnetization, AC susceptibility, and resistivity show no anomalous changes in near 80 K. Considering the ferromagnetic-like transition at 10 K shown in magnetic susceptibility measurements and the antiferromagnetic Weiss field at high temperature, we postulate that the ferromagnetic ordering at 78 K in KCo$_{2}$Se$_{2}$ is suppressed down to 10 K for CoSe.

\par The suppressed ordering may arise from geometric frustration, vacancies on the Co sites, or competing interactions between magnetic Co$^{2+}$ ions. In the case of CoSe, we can eliminate two of these possibilities: vacancy ordering and geometric frustration. Elemental analysis from previous work showed that the percentage of Co vacancies did not exceed 2\%, within error of that amount. Not enough to significantly suppress ordering. Geometric frustration occurs in systems where magnetic sublattices cannot arrange in a unique lowest energy ordered state, such as in an antiferromagnetic triangular lattice. CoSe contains a square lattice of cobalt atoms that cannot host this type of geometric frustration. Theoretical work, however, on square lattices have found frustration when the nearest neighbor and next-nearest neighbor magnetic interactions compete. This has been termed interaction frustration.\cite{vannimenus1977theory, shannon2004finite}

\subsection{Anisotropy and Magnetic Direction}

\par Our previous results from powder neutron diffraction indicated that the magnetic moments are aligned along the $c$-axis, contrary to the ordering in the related ``122''-phase.\cite{ZhouJACS} A possible reason for a difference in moment direction between the two systems could be could be due to closer CoSe layers  in the CoSe than in KCo$_{2}$Se$_{2}$. To understand the anisotropy present in the system, we performed single crystal magnetization measurements similar to the work done on KCo$_{2}$Se$_{2}$ by Yang \textit{et. al}. 

\par
\begin{figure}[t!]
	\includegraphics[width=3in, height=6.5in]{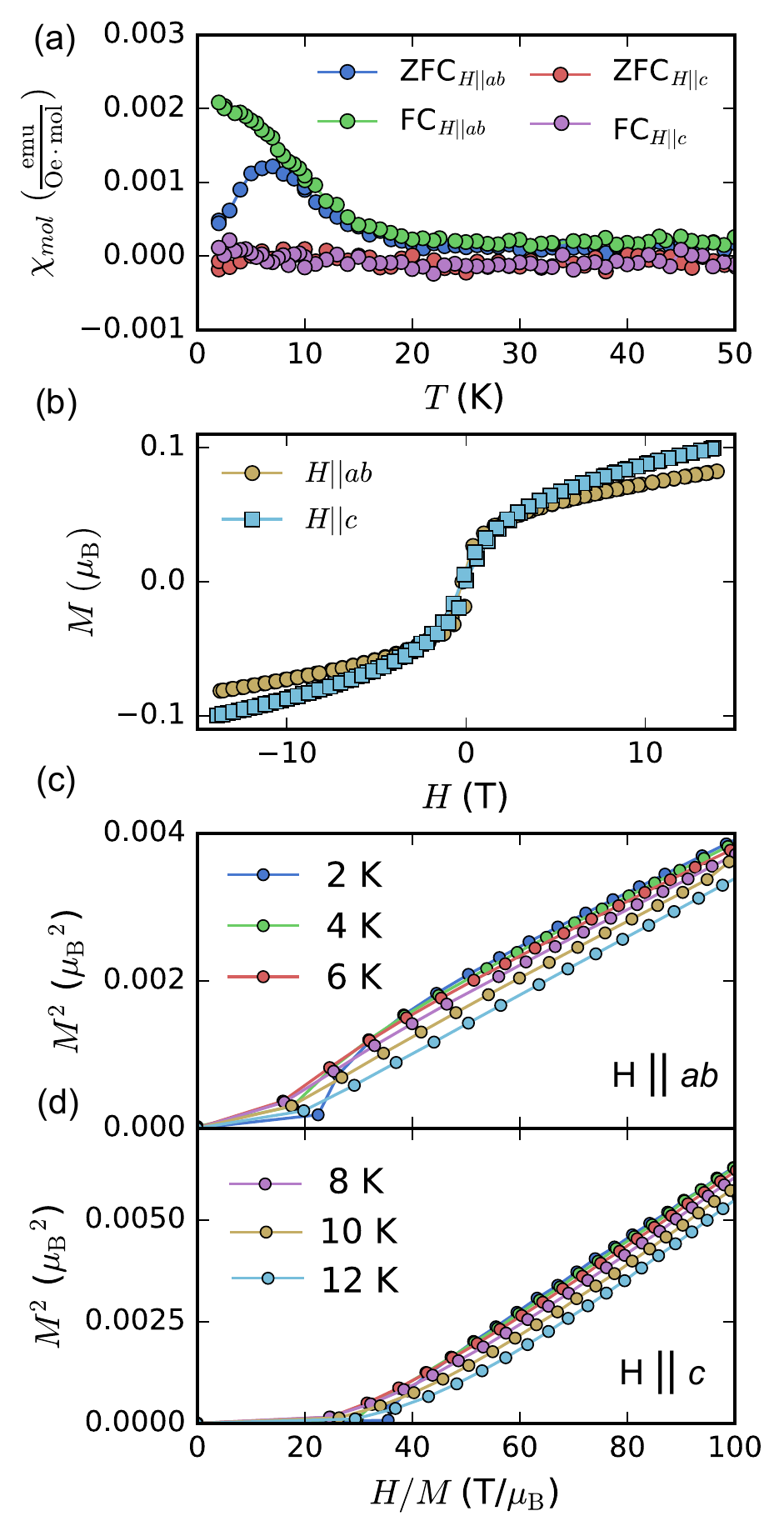}
 	\caption{Magnetic measurements of CoSe crystals mounted on a quartz paddle with orientations relative to the applied field direction as listed. a) Temperature dependent magnetic susceptibility for a 100 Oe field applied in two different orientations. b) Field dependent magnetization for both field orientations. c) Arrott plots constructed from $M(H)$ curves from 2 - 12 K for $H \parallel ab$, which indicate a ferromagnetic transiton within the 8 - 10 K range. d) Arrott plots constructed for $H \parallel c$-axis showing no spontaneous magnetization in the $c$-direction for any temperature.}
  	\label{figure:CoSe_anisotropy}
\end{figure}

 In Figure \ref{figure:CoSe_anisotropy}a, we see that magnetic susceptibility in the $ab$-direction is about four to five times larger than the susceptibility in the $c$-direction. This suggests a fair amount of anisotropy, but not as large as in KCo$_{2}$Se$_{2}$, where there is an a order of magnitude difference between the two field directions.\cite{yang2013magnetic} Unexpectedly, the anisotropy in the field dependence of the magnetization for KCo$_{2}$Se$_{2}$ did not hold for CoSe (Figure \ref{figure:CoSe_anisotropy}b). We see that for both field directions the magnetization does not saturate up to 14 T and approaches a moment value of 0.1 $\mu_{B}$. The itinerant nature of the magnetism leads to an unsaturated magnetization.

%-----------------------------------------------------------------------------------
%-----------------------------------------------------------------------------------

\par An important insight from these measurements is obtained by Arrott plot analysis. From Landau theory, we can expand the free energy, $F(H, T, M)$, of a magnetic system in the order parameter, $M$,  corresponding to magnetization: $F = F_{0} + aM^{2} + bM^{4}$. Minimizing the free energy with respect to magnetization we arrive at:\cite{mohn2006magnetism}

\begin{equation}
 M^{2}(T,H) = \frac{1}{b}\bigg(\frac{H}{M(T,H)}\bigg) - \frac{a}{b}\bigg(\frac{T-T_{c}}{T_{c}}\bigg)
 \label{eqn:Arrott}
\end{equation}

\par We can plot $M^{2}$ vs. $H/M$, known as an Arrott plot, to obtain linear relationships between $M^{2}(H)$ curves at different set temperatures. From Equation \ref{eqn:Arrott}, as the temperature approaches $T_c$, the $M^{2}(H)$ curves approach zero. Positive $y$-intercept values correspond to spontaneous magnetization at those temperatures. From Figure \ref{figure:CoSe_anisotropy}c for $H \parallel ab$-plane, the critical temperature appears to be in the 8 - 10 K range, as linear extrapolations of the $M^{2}(H)$ curves yield a zero y-intercept between 8 K and 10 K, which corresponds to transition temperature in the powder measurement. However, for $H \parallel c$-axis (Figure \ref{figure:CoSe_anisotropy}d), no $M^{2}(H)$ curves yield positive extrapolations back to the $y$-axis. Therefore, no spontaneous ferromagnetic moment orders along with $c$-axis.

\par The Arrott plot analysis matches previous reports for KCo$_{2}$Se$_{2}$, where the moment is claimed to lie solely in the $ab$-plane.\cite{yang2013magnetic} However, what causes the difference both the in ordering temperature and strength of the ferromagnetism between the two systems? The removal of potassium ions between the layers affects a number of factors: 1) cobalt oxidizes from Co$^{1.5+}$ in KCo$_{2}$Se$_{2}$ to Co$^{2+}$ which means a removal of electron carriers, 2) CoSe layer distances are reduced from 6.92 $\text{\AA}$ in KCo$_{2}$Se$_{2}$  to 5.33 $\text{\AA}$ in CoSe which may cause more effective exchange between the moments in adjacent $ab$-planes, and 3) the Co-Co distance shrinks from 2.710(3) $\text{\AA}$ in KCo$_{2}$Se$_{2}$ to 2.6284(3) $\text{\AA}$  in CoSe, which causes more orbital overlap between Co centers.

\subsection{FeSe vs. CoSe}

Currently, Fe and Co are the only transition metals that have been able to form the anti-PbO structure which is closely related to the parent ThCr$_{2}$Si$_{2}$ structure. The ThCr$_{2}$Si$_{2}$ hosts over 1,500 structures and a wide-range of physical phenomena. The anti-PbO phases are structurally simpler and can be used as the building blocks to systematically explore the physics within this structure type and, in general, metal square lattices. \cite{TTMC} 

\par Unconventional superconductivity has emerged in the Fe$Ch$ systems with the pairing mechanism for this phenomena still to be understood. With the close proximity of magnetism and superconductivity in the iron system, we need to understand the salient differences between CoSe and FeSe. Previous work directly compared the band structures of FeSe and CoSe and showed they differed by just a rigid band shift corresponding to the extra electron added by cobalt as compared to iron. \cite{ZhouJACS} This shift moved the Fermi level away from the nesting of hole and electron pockets evident in the FeSe superconductor, which could to be key to realizing superconductivity in this system.

\par Since band structure measurements have yet to be conducted on CoSe, we can directly compare the results of recent studies on KCo$_{2}$Se$_{2}$ and AFe$_{y}$Se$_{2}$.\cite{liu2012three, Qian_KFe2Se2} ARPES studies have shown that going from AFe$_{y}$Se$_{2}$ to KCo$_{2}$Se$_{2}$ (i.e. electron charge doping) changes the $3d$ orbital that contributes the most at the Fermi level. ARPES work on the AFe$_{y}$Se$_{2}$ series showed that the $3d_{xy}$ orbitals contribute the most at the Fermi level. The Se $4p_{z}$ orbitals also contribute to allow superexchange interactions. However, for KCo$_{2}$Se$_{2}$ the most significant orbital is the $3d_{x^{2} - y^{2}}$ which would change the interactions allowed between adjacent Co atoms.\cite{liu2015orbital} This change in geometry of the $d$-orbital likely is the mechanism for tuning away superconductivity to frustrated magnetism in CoSe.

\par Extensive work has been performed to understand the magnetic fluctuations in FeSe which are integral in understanding the mechanism responsible for superconductivity in the iron-based superconductors. As previously stated, the interesting interplay of magnetism in this system seems to stem from the electronic instabilities that accompany the square lattice formation.\cite{TTMC} Recent inelastic neutron diffraction work and theoretical work has shown that within the FeSe layers there is strong frustration between different magnetic ordered states (stripe \textit{vs.} N\'eel), which causes FeSe to not exhibit a true long-range magnetically order state.\cite{wang2016magnetic, wang2015nematicity, glasbrenner2015effect} The magnetic ordering in CoSe appears to suffer from similar frustration via the square lattice motif, although single crystal inelastic neutron spectroscopy measurements would shed further light on this hypothesis.

%%%%%%%%%%%%%%%%%%%%%%%%%%%%%%%%%%%%%%%%%%%%%%%%%%%%%%%%%%%%%%%%%%%%%
%% Conclusion
%%%%%%%%%%%%%%%%%%%%%%%%%%%%%%%%%%%%%%%%%%%%%%%%%%%%%%%%%%%%%%%%%%%%%
\section{Conclusion}

The synthesis of isostructural CoSe has allowed extensive characterization of the magnetic and transport properties of the system to understand its proximity to the iron-based superconducting analogues. Magnetic measurements have shown a transition reminiscent of ferromagnetism at 10 K with low applied fields that is fully suppressed at high fields. AC-susceptibility shows non-zero out-of-phase contributions, and such time dissipative magnetization below 10 K is indicative of a spin glass. Our more detailed analysis of the AC-susceptibility matches the behavior of CoSe to a spin glass, and we a possible explanation is the physics of interaction frustration present in square lattices.

\par Our Arrott plot analysis of the magnetization data reveals that the moment in CoSe lies within the $ab$-plane much like in related KCo$_{2}$Se$_{2}$. However, even if these two systems have similar anisotropy, the transition temperature is vastly different, having been suppressed from 78 K to 10 K in CoSe. Therefore, the amount of electron doping and density of states at the Fermi level can be used to tune the magnetic interactions in the Co square sublattice.

\par Resistivity measurements indicate a metallic state in CoSe with no significant anisotropic magnetoresistance and no discontinuity at the 10 K transition. Heat capacity measurements indicate no observable transition at 10 K either, but low temperature analysis reveals an enhanced Sommerfeld coefficient due to strong spin fluctuations at low temperatures. The lack of a discernable transition within transport measurements further corroborates the glassy character at low temperatures due to interaction frustration. Comparing CoSe to FeSe, we now see that the nature of the $d$-orbital occupany near the Fermi level vastly tunes the ground state from a metal with weak and competing magnetic interactions (CoSe) to a superconductor (FeSe).

\par Future work on the CoSe system includes inelastic neutron spectroscopy to shed further light on the nature of the exchange interactions leading to interaction frustration. Chemical manipulation to charge dope CoSe would also be an important step in further expanding the phase diagram of these metal square lattices. There has been some previous cobalt doping studies on FeSe but the amount of substitution on cobalt has been limited to less than 20\% due to phase stability with increased cobalt content.\cite{Mizuguchi_CoDope, urata2016non} However, the topochemical de-intercalation route should be able to expand the solid solution of cobalt-doped FeSe available to directly observe how superconductivity evolves into frustrated magnetism.

%%%%%%%%%%%%%%%%%%%%%%%%%%%%%%%%%%%%%%%%%%%%%%%%%%%%%%%%%%%%%%%%%%%%%
%% Acknowledgments
%%%%%%%%%%%%%%%%%%%%%%%%%%%%%%%%%%%%%%%%%%%%%%%%%%%%%%%%%%%%%%%%%%%%%
\section{Acknowledgments}
 	
Research at the University of Maryland was supported by the NSF Career DMR-1455118, the AFOSR Grant No. FA9550-14-10332, and the Gordon and Betty Moore Foundation Grant
No. GBMF4419. We also acknowledge support from the Center for Nanophysics and Advanced Materials. The authors acknowledge the University of Maryland supercomputing
resources (http://www.it.umd.edu/hpcc) made available for conducting the research reported in this paper. A portion of this work was performed at the National High Magnetic Field Laboratory, which is supported by National Science Foundation Cooperative Agreement No. DMR-1157490 and the State of Florida

%\bibliography{CoSe_Bib}
%\bibliographystyle{apsrev4-1}

%merlin.mbs apsrev4-1.bst 2010-07-25 4.21a (PWD, AO, DPC) hacked
%Control: key (0)
%Control: author (72) initials jnrlst
%Control: editor formatted (1) identically to author
%Control: production of article title (-1) disabled
%Control: page (0) single
%Control: year (1) truncated
%Control: production of eprint (0) enabled
%

\pagebreak
\widetext
\begin{center}
\textbf{\large Supplemental Materials: Frustrated magnetism in tetragonal CoSe, analogue to superconducting FeSe}
\end{center}
%%%%%%%%%% Merge with supplemental materials %%%%%%%%%%
%%%%%%%%%% Prefix a "S" to all equations, figures, tables and reset the counter %%%%%%%%%%
\setcounter{equation}{0}
\setcounter{figure}{0}
\setcounter{table}{0}
\setcounter{page}{1}
\makeatletter
\renewcommand{\theequation}{S\arabic{equation}}
\renewcommand{\thefigure}{S\arabic{figure}}
%%%%%%%%%% Prefix a "S" to all equations, figures, tables and reset the counter %%%%%%%%%%

%-----------------------------------------------------------------------------------
\begin{figure}[h!]
	\includegraphics[width=3in, height=3in]{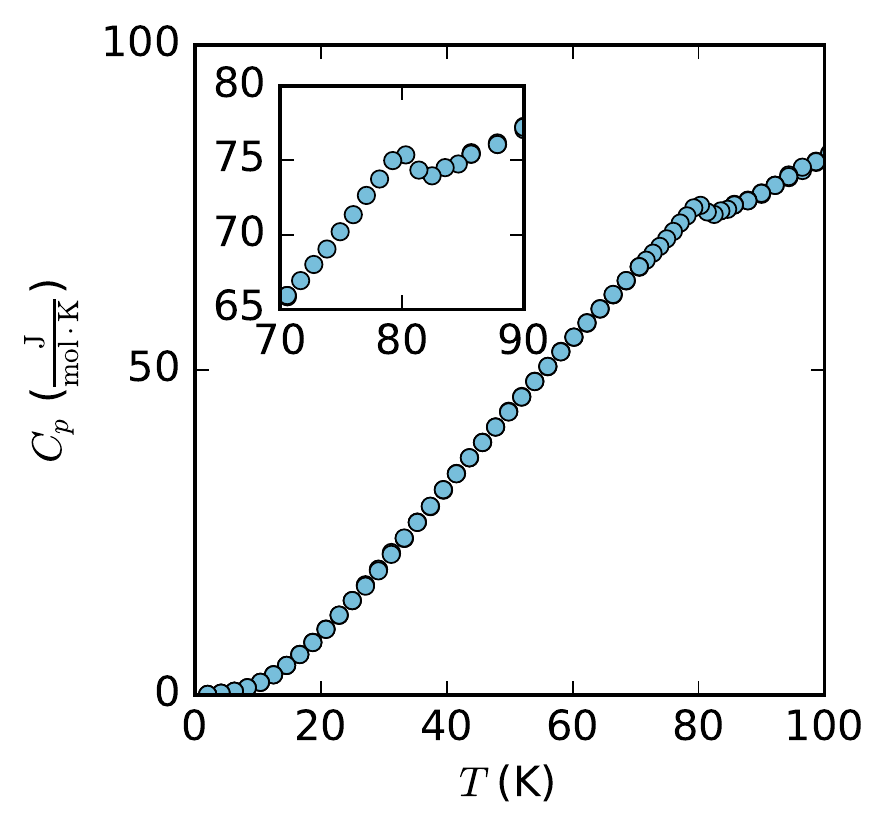}
 	\caption{Specific heat as a function of temperature measured on a single crystal of KCo$_{2}$Se$_{2}$ measured from 1.8 - 100 K with no applied field. Specific heat shows a distinct discontinuity at 79 K which corresponds to previous reports of the ferromagnetic transition in KCo$_{2}$Se$_{2}$; this is used to further prove the lack of KCo$_{2}$Se$_{2}$ phase within our CoSe samples signifying that complete de-intercalation is achievable for this system.}
  	\label{figure:KCo2Se2_heatcapacity}
\end{figure}
%-----------------------------------------------------------------------------------

All density functional theory (DFT) \cite{Hohenberg, Kohn} calculations were performed by using the Vienna Ab-initio Simulation Package (VASP)\cite{KresseThesis, KresseMD, KresseCalc, KresseIterative} software package with potentials using the projector augmented wave (PAW)\cite{BlochlPAW} method. The exchange and correlation functional were treated by the generalized gradient approximation (PBE-GGA).\cite{PerdewGradient}  The cut-off energy, 450 eV, was applied to the valance electronic wave functions expanded in a plane-wave basis set. A Monkhorst-Pack\cite{MonkhorstBrillouin} generated 23$\times$23$\times$17 k-point grid was used for the Brillouin-zone integration to obtain accurate electronic structures.

%-----------------------------------------------------------------------------------
\begin{figure}[h!]
	\includegraphics[width=4.5in, height=3in]{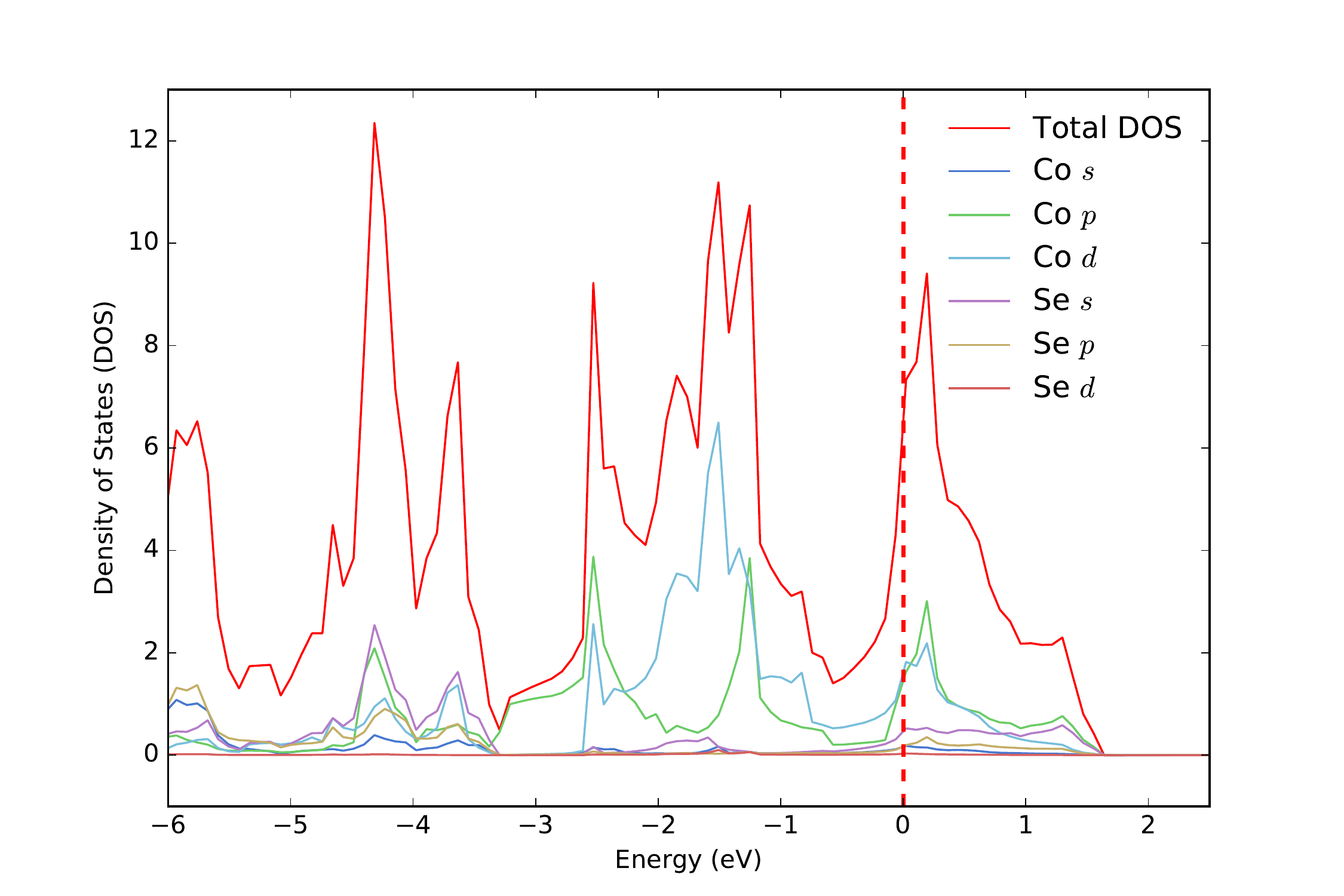}
 	\caption{Non-spin polarized density of states for CoSe decomposed for Co and Se atoms and their corresponding molecular orbitals.}
  	\label{figure:total_ldos_nonspin}
\end{figure}
%-----------------------------------------------------------------------------------

%-----------------------------------------------------------------------------------
\begin{figure}[h!]
	\includegraphics[width=4.5in, height=3in]{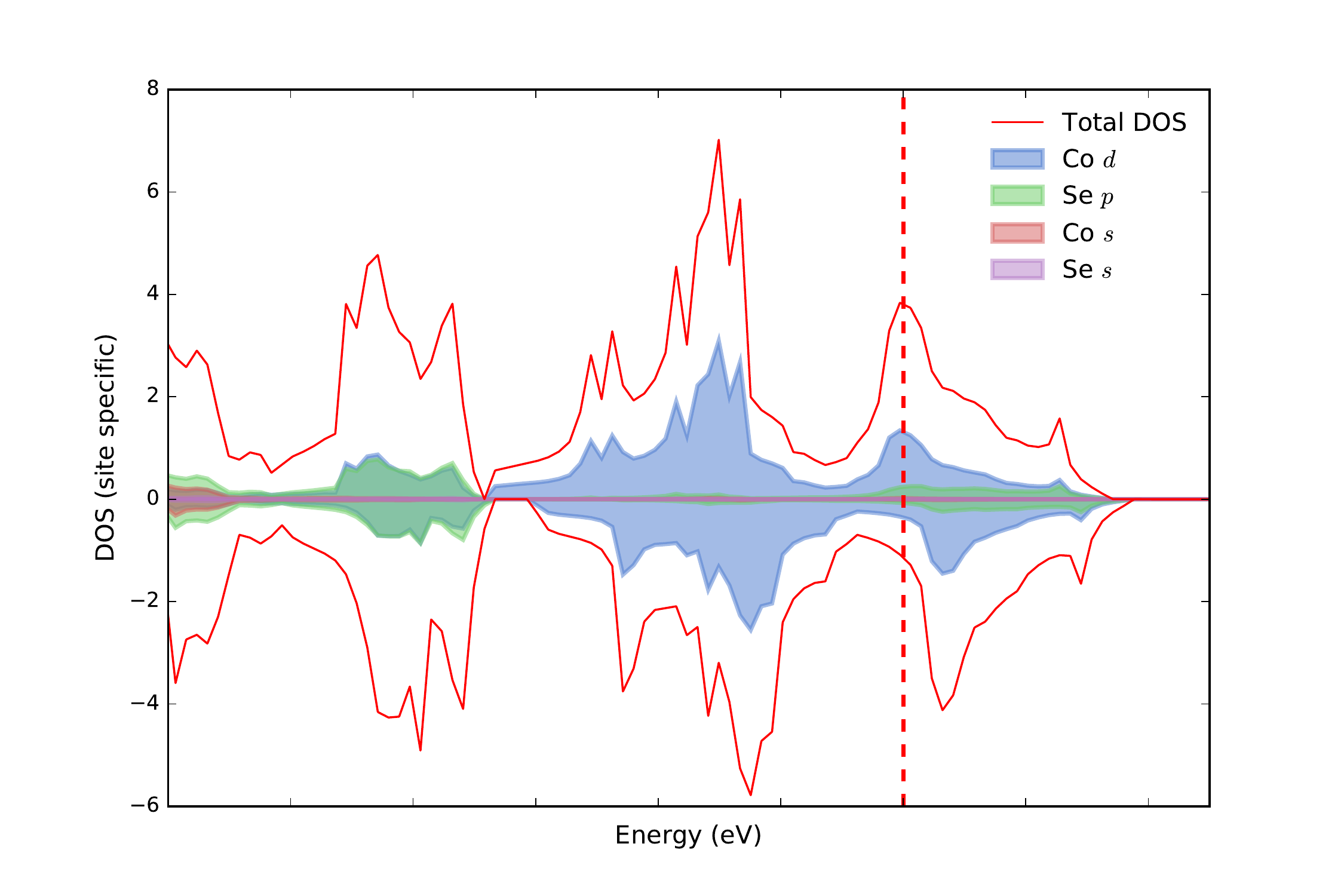}
 	\caption{Spin polarized density of states for CoSe decomposed for Co and Se atoms and their corresponding molecular orbitals.}
  	\label{figure:total_ldos}
\end{figure}
%-----------------------------------------------------------------------------------

%-----------------------------------------------------------------------------------
\begin{figure}[h!]
	\includegraphics[width=4.5in, height=3in]{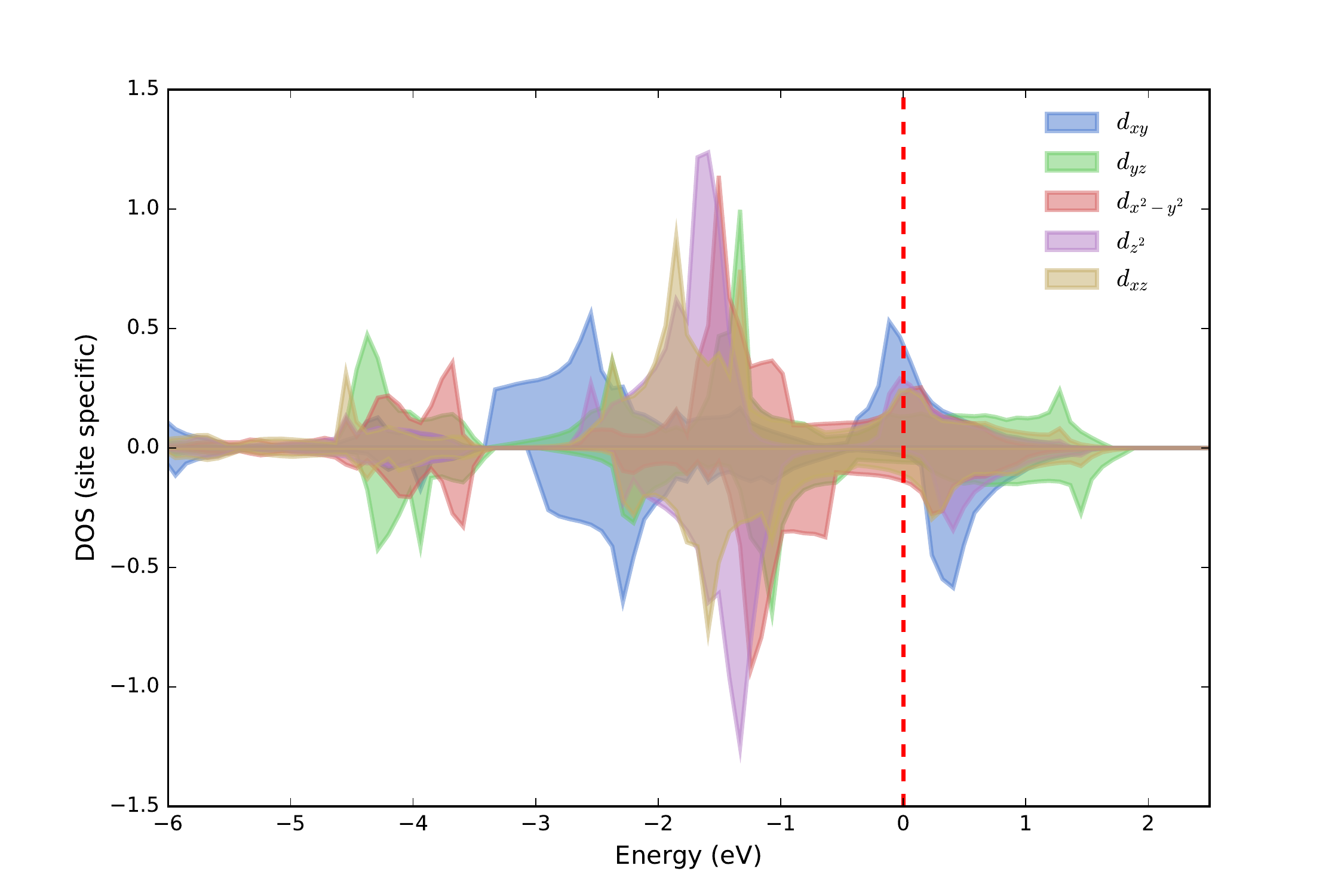}
 	\caption{Spin polarized density of states for CoSe decomposed for Co and the corresponding $d$-orbitals.}
  	\label{figure:Co_ldos}
\end{figure}
%-----------------------------------------------------------------------------------

%-----------------------------------------------------------------------------------
\begin{figure}[h!]
	\includegraphics[width=4.5in, height=3in]{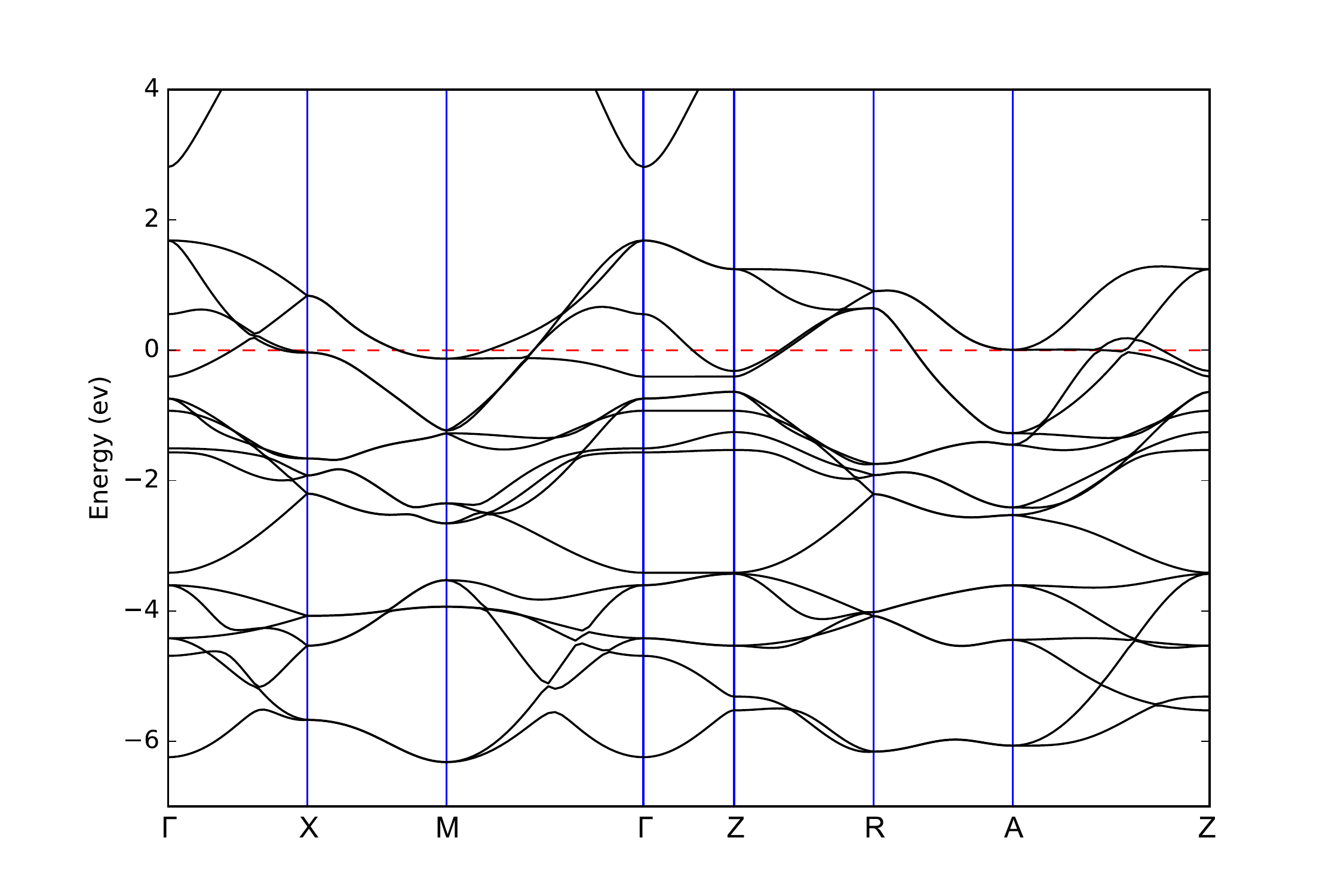}
 	\caption{Band structure for CoSe along high symmetry directions.}
  	\label{figure:bandstructure}
\end{figure}
%-----------------------------------------------------------------------------------

\end{document}